\documentstyle[proceedings,numreferences]{crckapb}

\def\lta{\;\rlap{\lower 2.5pt                       
             \hbox{$\sim$}}\raise 1.5pt\hbox{$<$}\;}

\def\msun{M_{\odot}}

\begin{opening}
\title{Millisecond Time Variations of X-Ray Binaries}
\author{J. H. Swank}
\institute{NASA/GSFC\\
Greenbelt,MD 20771, U.S.A.}
\end{opening}

\begin{document}


\begin{abstract}

Millisecond time-scales are natural for some neutron star and black hole
processes, although possibly difficult to observe. The Rossi X-Ray
Timing Explorer (RXTE) has found that for
the neutron stars in low-mass X-ray binaries (LMXB) there are flux
oscillations at high frequencies, with large amplitudes.
Z sources and bursters
tend to exhibit oscillations in the range 300-1200 Hz.
Persistent emission may exhibit one or both
of two features. In bursts from different bursters, a nearly coherent
pulsation is seen, which may be the rotation period of the neutron star.
For
some the frequency equals the difference between
the two higher frequencies, suggesting a beat frequency model, but in
others it is twice the difference.  The sources span two
orders
of
magnitude in accretion rate, yet the properties are similar. 
The similar maximum frequencies suggests that it corresponds to the Kepler
orbit frequency at the
minimum stable orbit or the neutron star surface, either of which would
determine
the
neutron star masses, radii and equation of state. Theories of
accretion onto black holes predict a
quasi-periodic oscillation (QPO) related to
the inner accretion disk. The two microquasar black hole candidates (BHCs)
have exhibited
candidates for this or related frequencies. 
 
\end{abstract}

\section*{Dynamical Time-scales of Compact Stars}

The Newtonian Keplerian frequency in an orbit at radius R about a compact
mass M
is a measure within a
factor of a few of the time scales of various other phenomena, such as
free fall, or oscillations, that can occur closest to the compact object.
For a $1.4 M_{\odot}$ neutron star with a 10 km
radius, the frequency is 2500 Hz, corresponding to a period of 0.4 ms. For
a $16 M_{\odot}$ black hole and $R = \frac{6GM}{c^2}$, the frequency would be
127 Hz with a 7 ms period. The effective area of the Proportional Counter
Array (PCA) (0.7\ m$^{2}$), the time resolution ($1 \mu$s), the data
modes, and the telemetry ($30-512$ kps are possible), were all required
to
study this regime.

It is clear from the results of RXTE 
that 
dynamical phenomena close to neutron stars and stellar black holes do 
generate X-ray signals which vary on the dynamical time-scales and that, 
furthermore, the signals are not greatly smeared by scattering  before 
escaping. Thus we are able to see oscillations of various sorts which 
are in the expected frequency regimes. 

The signals 
that have been found for neutron stars in the frequency range 50 Hz to
1300 Hz, starting with the discovery in
4U 1728-34 \cite{Stroh96}, are very 
significant (in many cases more than 10 sigma).
The highest quasiperiodic signal that has been
reported is about 1220 Hz. Many of the QPO features 
are narrow; they have $Q = \frac{\Delta\nu}{\nu} = 100-1000$. 

During the first year of the RXTE mission, two BHCs  were very bright, 
GRO\ J1655-40 and GRS\ 1915+105. In both of
these, a feature which qualifies as related to the dynamical time-scale of
the inner accretion disk has been seen, 300 Hz and 67 Hz, respectively 
\cite{Remillard97}. While for
the LMXB QPO the fractional root mean square (rms) power of the QPO were
1-20\%, 
for these BHC features the fractional rms amplitudes  were less
than 1\%.

\section*{High Frequencies from Neutron Stars}

Three kinds of frequencies higher than previously known  have 
been discovered in neutron star systems with RXTE
observations. 
There are now at least 6 low-mass X-ray binary sources of Type I bursts
in which 
almost coherent oscillations have been seen during the bursts. There 
are at least 14 low-mass X-ray binaries in which oscillations at 
frequencies ranging from 300 Hz to 1220 Hz have been seen in the 
persistent emission \cite{vdk97}.

The pattern of oscillation behavior is similar in many bursts.  When the
data is folded on the period, the pulsed light curve is
sinusoidal. A
spot slightly hotter than the rest of the neutron star and rotating with
it in and out of view would generate such an oscillation. 
The fractional root
mean square amplitude of the oscillation can exceed 50\% at the beginning
of the burst and decreases as the burst flux rises, as would be consistent
with the model of a burning front spreading on the neutron star.
We think the periods (1.7-2.7 ms)
probably indicate the
pulsars in these LMXB\cite{Stroh97}.

The kilohertz oscillations in the persistent emission from LMXB 
present a simple spectrum
in comparison to lower frequency 
power spectral densities (PSDs) of these sources.
Just two frequencies dominate over 
other signals.
Changes in the source luminosity usually
cause frequency change and the two features move together with the
difference approximately constant.
An obvious candidate for the higher frequency is the Kepler frequency  
of matter 
orbiting the neutron star  at the inner edge of the accretion disk. 
Various mechanisms could define the inner edge of the disk.
It could be (See \cite{Miller98} for extended discussion.)
interaction with the
magnetosphere, a sonic radius, or the marginally stable
orbit, if the  neutron 
star's surface is inside it.
A lower frequency would be seen at an 
alias with the neutron star rotation. If there is a clump circulating at 
the inner radius of the disk, accretion from it along field lines to the 
neutron stars magnetic pole would be most enhanced at the nearest 
conjunction of the pole and the clump.

\begin{table}
\begin{center}

\caption{LMXB with Kilohertz Oscillations}
\begin{tabular}{llllll}
\hline
Source &$L_{x}(10^{38})$
          &$Hz_{L}$ & $Hz_{H}$ &$\Delta$ & $Hz_{B}$   \\
\hline
4U 0614+09  &0.05 &326,400-800 &730-1145 &326    &    \\   
4U 1728-34  &0.06 &650-790     &500-1150 &363    &363 \\ 
Aql X-1     &0.08 &750         &         &       &549 \\ 
4U 1608-52  &0.1  &690,800-900 &900-1125 &60-230 &    \\
4U 1702-43  &0.1  &900         &1180     &280    &330 \\
4U 1636-53  &0.2  &800-900     &1170-1216&255    &581 \\
4U 1735-44  &0.4  &            &1149     &       &    \\
KS 1731-26  &0.2? &898         &1159-1207&260    &524 \\
X  1744-29? &0.2? &            &         &       &589 \\
4U 1820-30  &0.6  &546-796     &1066     &270    &    \\
GX 17+2     &1    &880-682     &988      &306    &    \\
Cyg X-2     &1    &490-530     &730-1020 &343    &    \\
Sco X-1     &1    &634         &926      &292    &    \\
            &2    &794-820     &1062-1133&247    &    \\
GX 5-1      &4    &325-448     &567-895  &325    &    \\
\hline
\end{tabular}
\end{center}
\end{table}

The sources with reported high frequency signals in the persistent
emission are summarized in Table 1 (luminosity in ergs s$^{-1}$,
frequencies in the
persistent emission(L and H), the difference($\Delta$), and in the
bursts(B)). It can
be seen that while for
4U 1728-34 the difference between the 2 high frequencies equals that of    
the bursts, for 4 other sources, the difference is approximately half the 
frequency seen during the bursts. However the burst period does not appear
interpretable as the harmonic of the spin because of the high pulsed
amplitude and the sinusoidal form.

The maximum frequencies appear to be all very close to each other.
This suggested 
that the determination of the frequency must be dominated by neutron star
characteristics and roughly independent of the accretion rate and the
optical depths of the formation region. 
We argued that most likely the equation of state of 
nuclear matter does lead to neutron stars with radii inside the
inner-most 
stable orbit for a  neutron star, and the disk cannot approach closer
to the neutron star than this\cite{ZhangW97}. 
The observed frequencies of about 1200 Hz would imply neutron 
star masses of $1.8-2 \msun$. The existence 
of an orbit which is marginally stable would be a verification of an
important consequence of strong gravity\cite{Kaaret97}.

\section*{BHC oscillations}

The frequency of fast 
oscillations in the inner disk around a black hole depends on the black
hole mass, its 
angular momentum, and the model. Instabilities of the disk which lead
to 
circulating clumps would imply the masses of 7 and 33 $\msun$ for 
GRO\ J1655-40 and GRS\ 1915+105, respectively, if they are slowly
rotating. 
Epicyclic frequencies would match those observed for smaller slow black 
holes, but for the more massive black holes if they have near maximum 
rotation rate\cite{Nowak97}. The mass and the angular momentum of the hole
imply the 
inner radius of the optically thick accretion disk and the inner radius
is more consistent with the mass of GRO J1655-4
deduced 
from optical measurements  if the hole is rotating fast\cite{ZhangN97}. 


\end{document}